\documentclass[12pt,preprint]{aastex}

\shorttitle{2002 Outburst of 4U 1543-47}
\shortauthors{Buxton et al.}

\begin{document}


\title{The 2002 Outburst of the Black-Hole X-ray Binary 4U 1543-47: Optical and Infrared Light Curves}

\author{M.M. Buxton\altaffilmark{1}}
\email{buxton@astro.yale.edu}

\and

\author{C.D. Bailyn\altaffilmark{1}}
\email{bailyn@astro.yale.edu}

\altaffiltext{1}{Astronomy Department, Yale University, P.O. Box 208101, New Haven, CT, 06520-8101, U.S.A.}


\begin{abstract}
We have obtained simultaneous optical and near infrared observations of 4U 1543-47 during its 2002 outburst.  The most striking feature of the outburst light curve is the secondary maximum which appears after the object transitions into the low-hard state.  This secondary maximum is much stronger in the infrared bands than optical.  We suggest that the origin of the secondary maximum flux may be synchrotron radiation associated with a jet.  Close infrared monitoring may lead to reliable triggers for simultaneous multiwavelength campaigns to study jet formation processes.  
\end{abstract}




\section{Introduction}

Soft X-ray transients (SXTs) are a subclass of X-ray binaries which comprise a compact object (black hole or neutron star) and a low-mass ($\le$ 1 M$_{\odot}$) secondary star.  SXTs spend most of their time in quiescence but occasionally undergo X-ray outbursts during which the mass accretion rate through the disk increases dramatically and matter is accreted by the compact object.  These objects provide an excellent means of studying accretion processes and how they relate to the fundamental binary parameters 
that can be measured in quiescence.  Jets have been observed in SXTs, particularly during low-hard states \citep{fen01}.  SXTs are excellent candidates to answer outstanding questions about jet formation and behavior.  To do so, however, one must obtain simultaneous (or quasi-simultaneous) multiwavelength data in order to relate the timing and origin of emission from various parts of the binary.

The SXT, 4U 1543-47, was discovered on 17 August 1971 by the \textit{Uhuru} satellite \citep{mat72}.  Since then additional X-ray outbursts have been observed in 1983, 1992 and 2002.  The optical counterpart was found during the 1983 outburst by \citet{ped83}.  Dynamical studies of the companion star \citep{oro98} show that the compact object is most likely a black hole with a mass of 2.7 $\le$ M$_1 \le$ 7.5 M$_{\odot}$. 

\begin{figure}
\begin{center}
   \plotone{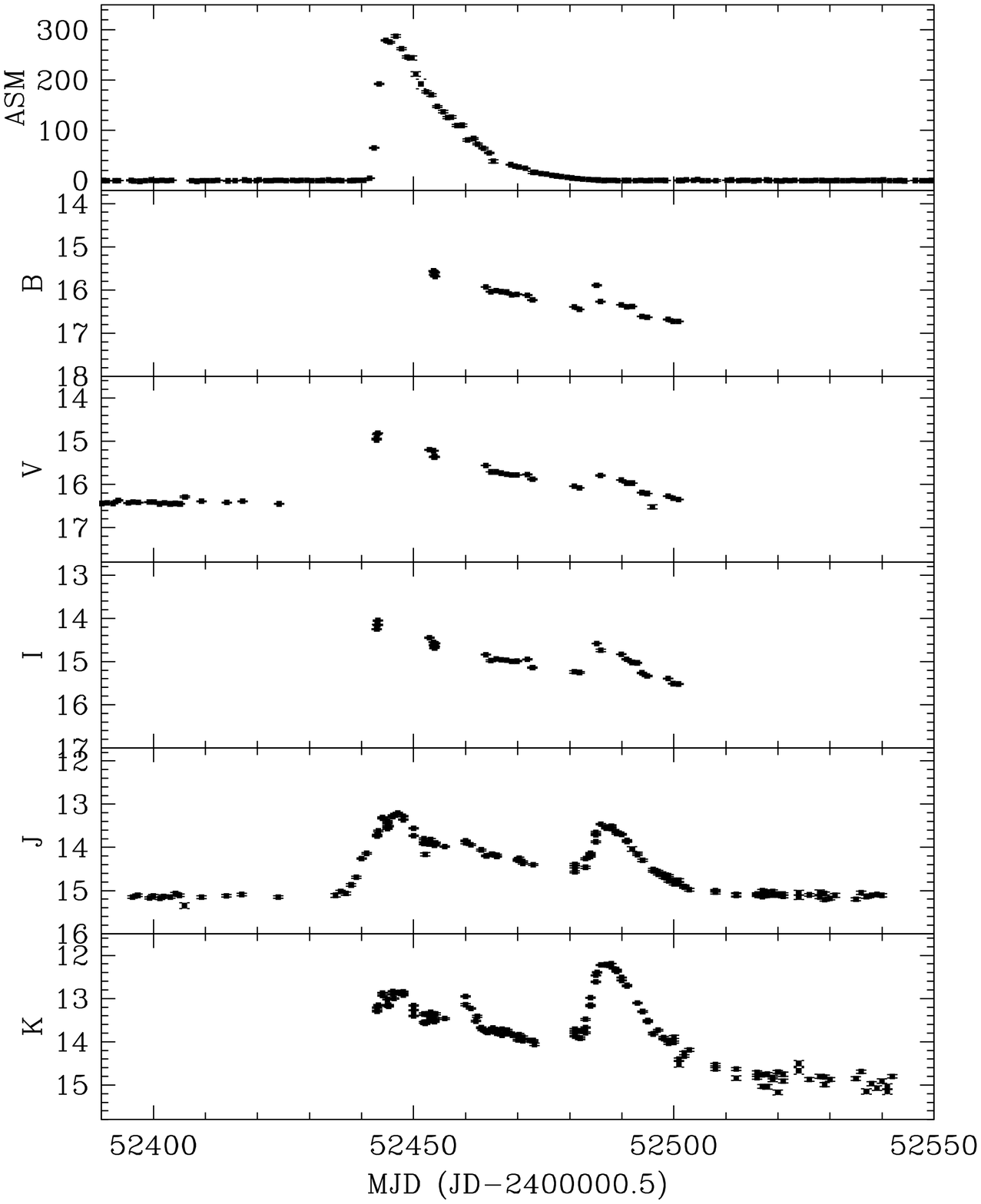}
    \figcaption{\small{Optical and infrared data of 4U 1543-47 in 2002.  The top panel is the one-day averaged RXTE ASM data provided by the ASM/RXTE team. }\label{fig:1543_alldata}}\end{center}
\end{figure}

The 2002 outburst began on MJD 52442 (UT June 16.683) as observed by RXTE \citep{mil02,par03}.  During this outburst the soft X-ray light curve exhibited a classical fast rise, exponential decay profile as observed in many other SXTs \citep{che97}.  The top panel in Figure \ref{fig:1543_alldata} shows the RXTE/ASM light curve from MJD 52390 to 52550.  At MJD 52443.5 the Molonglo Observatory Synthesis Telescope (MOST) detected a radio counterpart at 843 MHz with a flux density of 11 $\pm$ 1.5 mJy \citep{hun02}.  From early on in the outburst until around MJD 52477, 4U 1543-47 was in a high-soft state.  A transition then occurred on this day into the low-hard state \citep{kal02}.  
Radio observations late in the outburst are also reported by \cite{par03}.

We observed the 2002 outburst of 4U 1543-47 as part of the YALO consortium \citep{bai99}\footnote{Now operating as the Small and Medium Aperture Research Telescope System (SMARTS), www.astro.yalo.edu/smarts/}.  Our observations cover most of the outburst cycle in the infrared and some in the optical.  A ``secondary maximum'' appeared in the optical and infrared (OIR) light curves which is not apparent in the RXTE/ASM data but, near its peak, corresponds to the second MOST radio detection.  In this paper we present our OIR observations and reduction (\S \ref{sec:obs}), discuss the OIR light curve morphologies (\S \ref{sec:morphology}) and focus our analysis on the secondary maximum (\S \ref{sec:SED}).  In \S \ref{sec:discussion} we discuss our results and argue that the secondary maximum flux may be synchrotron emission which originates from a jet.  We also note that secondary maxima may be a common phenomenon in SXTs during outburst decline after the source transitions into the low-hard state.  In \S \ref{sec:conclusion} we give our conclusions.

\section{Observations, Data Reduction and Calibration}
\label{sec:obs}

$V$- and $J$-band images of 4U 1543-47 were obtained on a daily basis (weather permitting) from MJD 52323 (UT 2002 February 17) using the YALO 1.0m telescope together with ANDICAM\footnote{www.astronomy.ohio-state.edu/ANDICAM/}, a dual-channel imager capable of obtaining optical and infrared data simultaneously.  Data was recorded by a Lick/Loral-3 2048x2048 CCD on the optical channel and a Rockwell 1024x1024 HgCdTe "Hawaii" Array on the infrared channel.  On UT 2002 June 5 the YALO optical CCD failed.  However, $J$-band observations continued on a daily basis.  

When the $J$-band light curve was observed to have risen by $\sim$ 0.5 mag ($\sim$ MJD 52440) we initiated additional observations in the $K$-band.  Optical observations were obtained on a daily basis, when possible, from UT 2002 June 18 to August 15 (MJD 52442.9 - 52500.8) using the 74 inch telescope at Mount Stromlo Observatory with the Cassegrain imager, a 2K x 4K CCD and $B, V$ and $I$ filters.  Observations ceased when the object was no longer high enough in the night sky to observe.  

Exposure times for the optical frames were 300 sec for YALO images and 120 sec for Stromlo images.  For the infrared, 7 images of 90 sec each were obtained in $K$-band and 5 images of 60 sec in $J$-band.  Each image was shifted via an internal mirror in right ascension or declination by $\sim$40 arcsec.  

Optical data reduction, including bias subtraction and flat fielding, was performed using the usual routines in {\tt IRAF}.  Photometry was performed using {\tt DAOPHOT} in {\tt IRAF}.  Calibration was derived by calculating the offsets from a secondary star, the magnitude of which was measured previously (J. Orosz, private communication).  The magnitudes were corrected for airmass extinction using coefficients taken from the CTIO table in {\tt IRAF}.  

Infrared images were reduced using an in-house {\tt IRAF} script which flat fields (using dome flats), subtracts scaled sky images, shifts the images to a reference image then combines all images by averaging them.  Aperture photometry was performed using \textit{qphot} in {\tt IRAF}.  Calibration was done using the primary standard P9187\footnote{See Persson faint IR standards at\\www.ctio.noao.edu/instruments/ir\_instruments/ir\_standards/hst.html}, observed on UT 2002 September 2.  Magnitudes were corrected for airmass extinction using atmospheric extinction coefficients taken from \cite{fro98}.  

\section{Light Curve Morphology}
\label{sec:morphology}

Figure \ref{fig:1543_alldata} shows $B-, V-, I-, J-$ and $K$-band light curves together with the RXTE/ASM light curve.  Due to the lack of data in the optical bands it is more difficult to make qualitative or quantitative comparisons to the X-ray light curve.  Therefore, most of our analysis of the outburst rise and peak deals with the IR bands.  

\subsection{Start of Outburst}
\label{sec:outburst_start}

\begin{figure}
\begin{center}
   \plotone{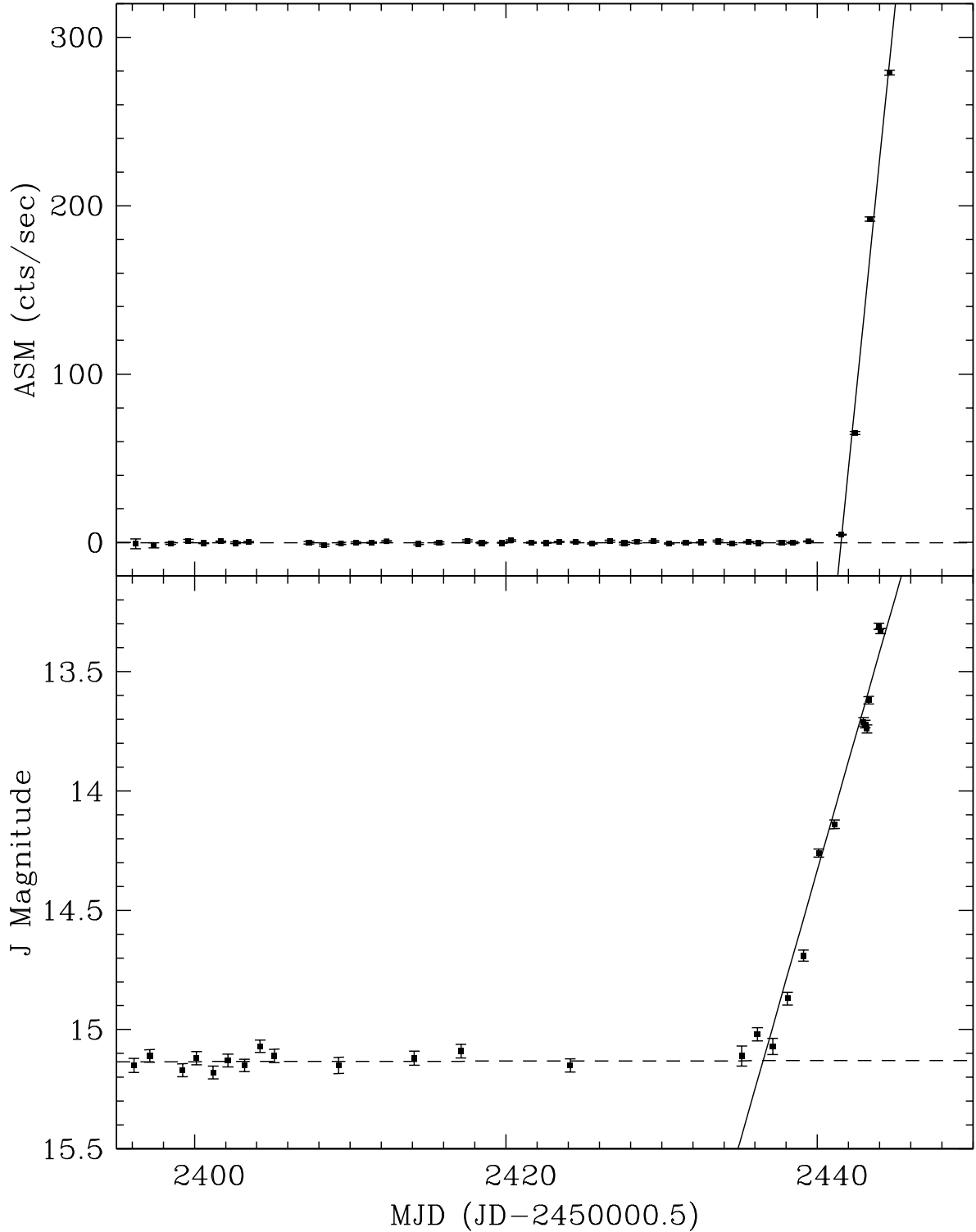}
    \figcaption{Quiescent period and early part of the 2002 X-ray outburst of 4U 1543-47.  The top panel is the one-day averaged RXTE/ASM data, bottom panel is $J$-band data.  The dashed line is the mean quiescent value.  The solid line is the fit to the outburst data.  The $J$-band data starts to rise $\sim$ 3-5 days before the X-rays.  \label{fig:rise}}
\end{center}
\end{figure}

The $J$-band light curve rises 
well before the RXTE/ASM flux becomes observable.  
In Figure \ref{fig:rise} we have plotted the RXTE/ASM data (top panel) and $J$-band data (bottom panel) for the period leading up to, and just after, the start of the outburst.  The mean value of the quiescent data was found to be 15.13 mag for $J$-band and 0 counts/sec for the ASM.  These are indicated by the dashed lines shown in Figure \ref{fig:rise}.  To determine the start of the outburst 
we fitted the outburst part of the light curves with a linear least-squares fit.  For the X-rays we fitted data between MJD 52441.5 - 52444.7 (four points).  
For the $J$-band outburst data, two fits were made.  The first included data from MJD 52435.5 - 52444.5, the second fit (shown as the solid line in Figure \ref{fig:rise}) was performed on data from MJD 52437.6 - 52444.5.  Including the two points at MJD 52435.6 and 52436.6 had a significant impact on the gradient of the line.  This was the dominant error in determining the time of the start of the outburst.  

The $J$-band outburst began at MJD 52436.78 $\pm$ 0.58 (including the error mentioned above) whereas the X-ray outburst began at MJD 52441.51 $\pm$ 0.01.  
Fitting the X-ray data in log space resulted in an outburst start time of
MJD 52439.38.  Hence, the start of the X-ray outburst 
was delayed by $\sim$ 3-5 days with respect to the infrared.  Similar delays have been seen in other SXTs \citep{oro97}.  The slopes of the rise are 91.91 $\pm$ 0.01 counts/sec/day for the X-rays and -0.24 $\pm$ 0.01 mag/day for the $J$-band.  Since the IR began to rise before the ASM, this suggests that the outburst was triggered in the outer regions of the accretion disk.  Caution must be given, however, as the sensitivity level of the RXTE/ASM is orders of magnitude above the quiescent flux level for this source, so fine
details of the X-ray light curve in the early stages of the outburst are
undetectable.  

\subsection{Outburst Maxima and Flares}

\begin{figure}
\begin{center}
   \plotone{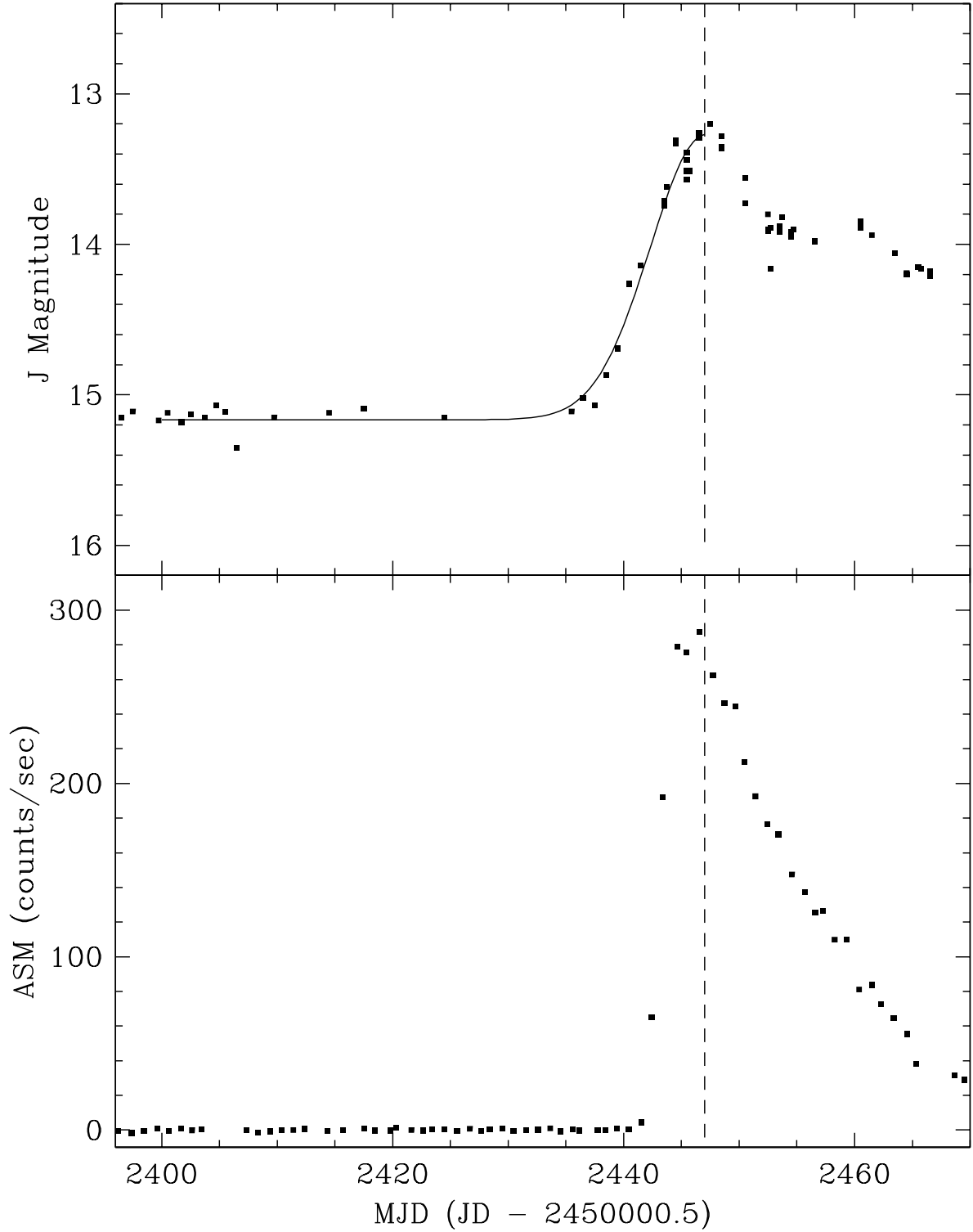}
   \figcaption{Gaussian fit to the initial rise in $J$-band light curve.  The vertical-dashed line indicates best fit peak time of MJD 52477.  The solid line is the rising portion of best-fit Gaussian.  \label{fig:gauss_rise_fit}}
\end{center}
\end{figure}

In comparing the $J$-band light curve to that of the ASM it is not apparent that there is any noticeable difference in the peak times.  We determined the peak time in the $J$-band light curve by fitting Gaussian profiles to the rising parts of the light curve.  We fixed the $y$-axis intercept in the fits to the mean values determined for the rise start times (\S\ref{sec:outburst_start}).  This was added as a constant to the Gaussian function.  We restricted the fit between MJD 52400-52450.  The peak, amplitude and Gaussian width were varied.  The best fit gave a peak time of MJD = 52447.1, amplitude = -1.9 mag and width = 4.8 days.  The best fit Gaussian is plotted in Figure \ref{fig:gauss_rise_fit}.  A fit was not performed on the X-ray data as the small number of datapoints gave large errors.  

After the primary X-ray peak, the X-rays and OIR decline until $\sim$ MJD 52460 when the $K$-band light curve, and to a lesser extent the $J$-band, experiences a small flare.  Just before this flare, the source transitioned briefly into a 
steep power-law state and low-frequency QPOs were observed by \cite{par03}.  No optical or infrared data were obtained by us during the time of transition.  At $\sim$ MJD 52462, the $K$-band returned to the previous decline rate from the primary outburst.  Since the overall decline was not interrupted, we suspect that the small flare emission is due to a mechanism different from that of the primary outburst.  

At $\sim$ MJD 52484 the $J$- and $K$-band fluxes experience a significant rise, which we call the ``secondary maximum''.  This rise in flux also appears in the optical bands but is much weaker.  The peak strength of the secondary maximum increases with wavelength, that is, it is smallest in $B$-band and highest in $K$-band.  

\begin{figure}
\begin{center}
   \plotone{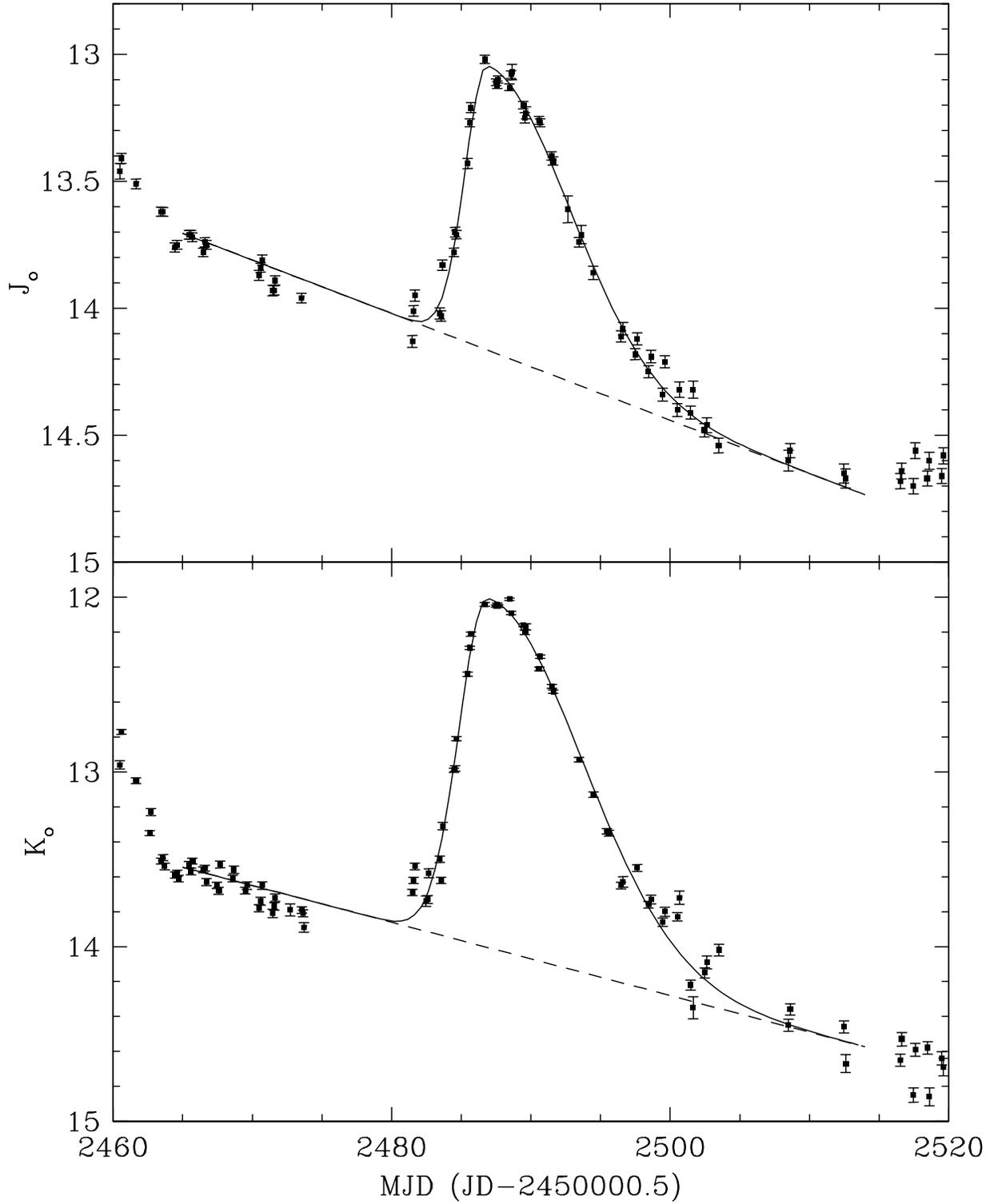}
    \figcaption{Two-sided Gaussian fits to the secondary maximium in $J$- and $K$-bands.  Dashed line is best-fit slope of underlying flux, solid line is best-fit two-sided Gaussian.  \label{fig:gauss_sec_max}}
\end{center}
\end{figure}

In Figure \ref{fig:gauss_sec_max} we have plotted a zoomed part of the $J$- and $K$-band light curves around the secondary maximum.  One can separate the two components:  a steadily declining flux from the primary outburst on top of which is superimposed additional flux from the secondary maximum.  The first component we will call the ``underlying flux''.  The secondary maximum is flux additional and separate to that of the linearly-decaying flux.  Hence, it seems likely that its origin is also separate from the processes governing the primary outburst.  

We estimated the peak times in each band by fitting two-sided Gaussians to the secondary maximum.  Two-sided Gaussians were used as they provided a much better fit than a single Gaussian.  The data were first detrended for the underlying flux by removing a linear component fitted to the data either side of the secondary maximum.  The best fit Gaussian parameters are shown in Table \ref{tab:sec_max_fits}.  The secondary maximum peaked in both wavebands at the same time:  in the $J$-band, the fitted peak time was MJD 52486.3$^{+1.5}_{-1.8}$ (1-$\sigma$) while for $K$-band it was MJD 52486.5$^{+2.1}_{-1.8}$ (1-$\sigma$).  

We reiterate here that the second MOST radio detection was made near the peak of the secondary maximum.   It is not known whether a radio source was present before or after this detection as observations were not conducted.  Although there is no obvious change in the RXTE/ASM light curve, there may be effects seen in the RXTE/PCA data which will be presented elsewhere (Kalemci et al., in prep.).  

After the secondary maximum, the OIR light curves return to the original decline rate of the primary maximum until reaching previous quiescent levels.
 
\section{Spectral Energy Distributions}
\label{sec:SED}

To understand the source of the secondary maximum emission, we have constructed a spectral energy distribution (SED) of the flux from the secondary maximum only at the time when the second MOST radio detection was made (MJD 52487).  The OIR magnitudes of 4U 1543-47 were first corrected for interstellar extinction using E($B-V$) = 0.5 $\pm$ 0.05 \citep{oro98}, A$_V$ = 3.2 E($B-V$) = 1.6 \citep{zom90}, A$_B$/A$_V$ = 1.324 and A$_I$/A$_V$ = 0.482, A$_J$/A$_V$ = 0.282 and A$_K$/A$_V$ = 0.112 \citep{rie85}.  As there were no $B, V$ or $I$ observations made on this date, we extrapolated the optical light curves to estimate their magnitudes.  For this date we estimate the secondary maximum magnitudes at the peak to be $B_o$=14.2, $V_o$=14.2 and $I_o$=13.9.  We estimated the magnitude of the underlying flux at the time of the secondary maximum peak to be $B_o$ = 14.5, $V_o$ = 14.6 and $I_o$ = 14.6.  In the IR bands, the secondary maximum peak is at $J_o$ = 13.1 and $K_o$ = 12.1 and the underlying flux magnitudes are $J_o$ = 14.3 and $K_o$ = 14.1.  All magnitudes were converted to fluxes using zeropoint fluxes given by \cite{all01} and \cite{cam85}.  In each band, the underlying flux was subtracted from the peak flux giving the residual flux, corresponding to the flux of the secondary maximum only. 

\begin{figure}
\begin{center}
   \plotone{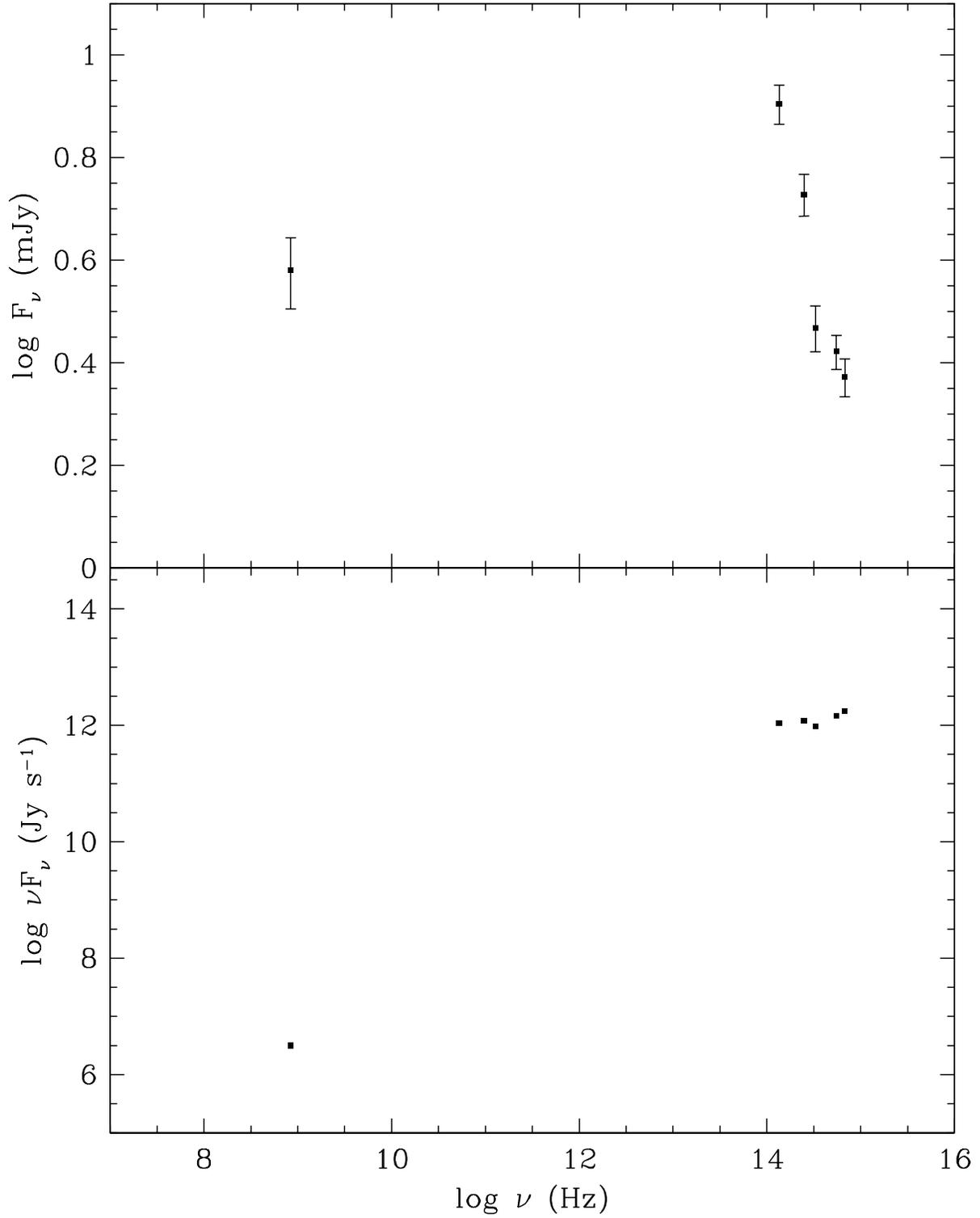}
    \figcaption{Spectral energy distribution of the secondary maximum flux on MJD 52487 (UT 2002 July 31).  The $B$, $V$ and $I$ points were extrapolated from their respective light curves.  \label{fig:sed}}
\end{center}
\end{figure}

The SED of the secondary maximum flux at its peak is presented in Figure \ref{fig:sed}.  The top panel shows the data in log F$_{\nu}$ v. log ${\nu}$ space and the bottom panel in log $\nu$F$_{\nu}$ v. log ${\nu}$ space.  In the top panel the optical and IR points lie almost on a straight line.  There is clearly a break somewhere between the $K$-band point and the radio but it is impossible to say where exactly.  In the bottom panel we can see that most of the power is in the OIR bands.  

\subsection{Blackbody fits}

\begin{figure}
\begin{center}
   \plotone{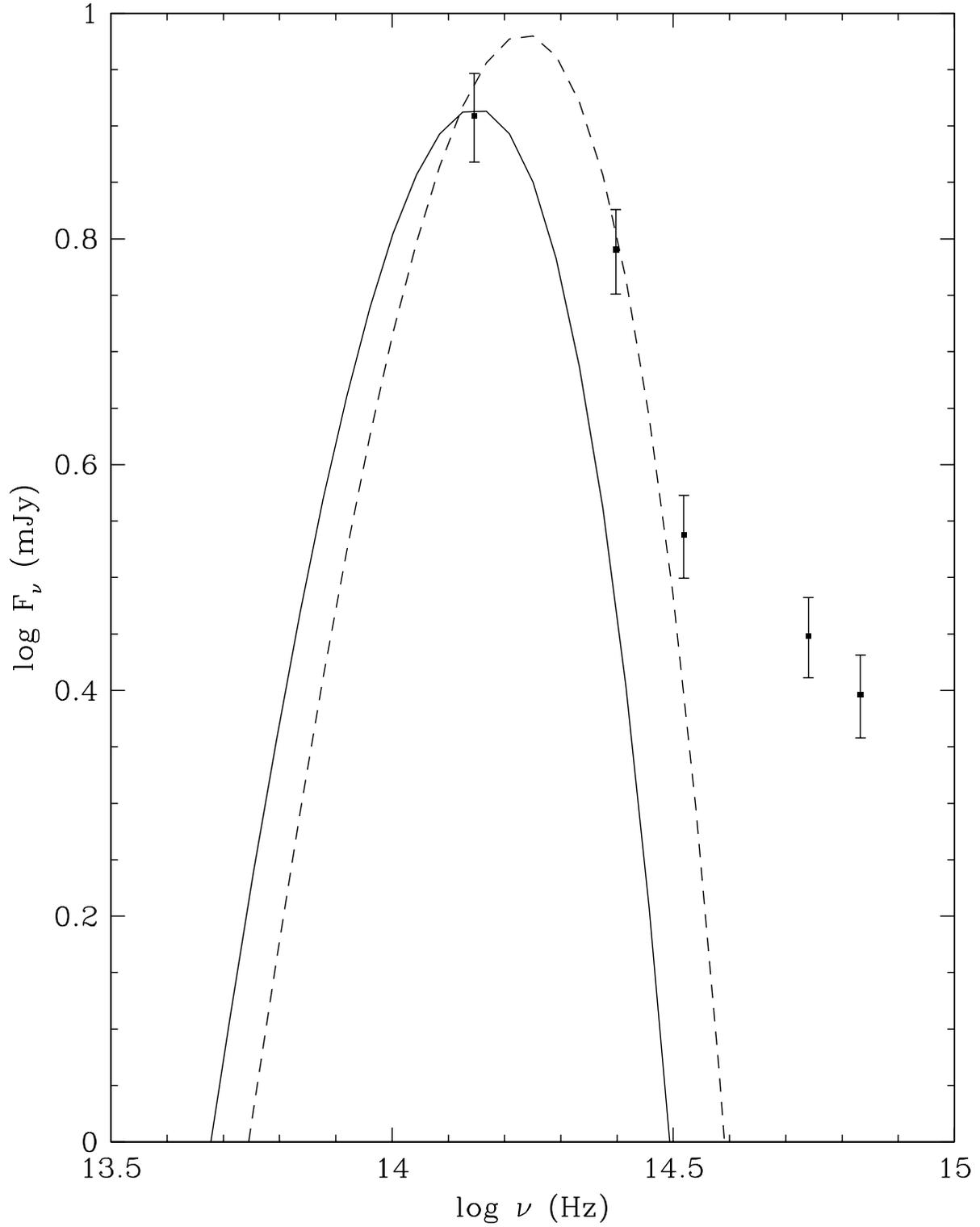}
    \figcaption{Single-temperature blackbody fits to the optical and infrared section of the SED of the secondary maximum flux.  Dashed line is for T$_{bb}$ = 1650$^o$K, solid line for T$_{bb}$ = 1350$^o$K.  These models cannot account for the optical/infrared flux observed.  \label{fig:sed_onebb} }
\end{center}
\end{figure}

We attempted to fit the SED with a single blackbody, peaking at or around the $K$-band point since this has the greatest flux.  Two example fits are shown in Figure \ref{fig:sed_onebb}.  
The solid line for  T$_{bb}$ = 1350$^o$K peaks at K, while
the dashed line is the best fit to the I, J and K data, with 
a blackbody temperature of T$_{bb}$ = 1650$^o$K.
Neither of these models fit the optical data.  
For a temperature of 1650$^o$K, the emitting region would need to have a radius of $\sim$ 1.4 x 10$^{17}$cm given a distance of 
9.1 kpc \citep{oro98}.  
For an orbital period of 1.123 days \citep{oro98}, the size of the primary Roche lobe is $\sim$ 5 x 10$^{11}$cm.  
Hence this immediately rules out an optically thick accretion disk as the source
of the observed flare radiation.
  
\begin{figure}
\begin{center}
   \plotone{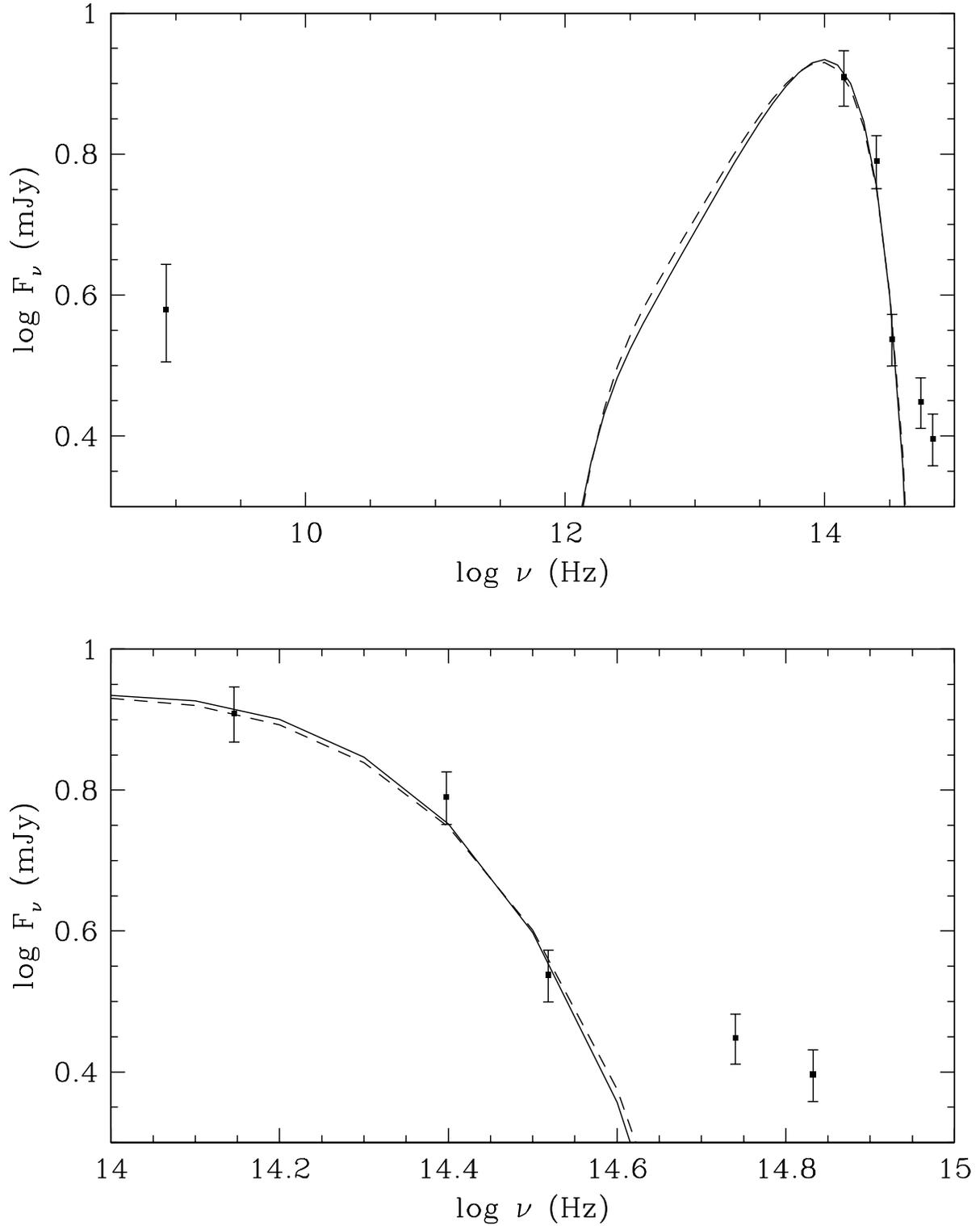}
    \figcaption{Two examples of multicolor-blackbody model fits to the SED of the secondary maximum flux.  See text for model parameters.  The multicolor-blackbody model is slightly better than the single-temperature blackbody model (Fig. \ref{fig:sed_onebb}) but there is still a large excess in the $B$ and $V$ bands, nor does it account for the radio flux. \label{fig:sed_multi} }
\end{center}
\end{figure}

The SED was fitted with multicolor-blackbody models \citep[see][for parameter definitions]{mit84}.  We fixed the inner radius, r$_{in}$ = 3r$_g$, corresponding to the innermost stable orbit of a black hole (where r$_g$ is the gravitational radius).  The mass of the compact object (black hole) was fixed at 5 M$_{\odot}$ \citep{oro98}.  In Figure \ref{fig:sed_multi} we show two fits: the first (solid line) is for a disk with r$_{out}$ = 10$^{7.04}$ r$_{in}$ (= 4.9 x 10$^{13}$cm) and $\dot{M}$ = 10$^{16}$ gs$^{-1}$, the second (dashed line) is for r$_{out}$ = 10$^{4.4}$ r$_{in}$ (= 1.1 x 10$^{11}$cm) and $\dot{M}$ = 10$^{8.3}$ gs$^{-1}$.  Although the models fit the $I$-, $J$- and $K$-band points, they cannot account for the $B$ and $V$ points.  In the first case, r$_{out} \sim$ 100 times the primary Roche lobe.  In the second case, r$_{out}$ is within the primary Roche lobe but $\dot{M}$ is far less than the average observed accretion rates of X-ray binaries, that is $\sim$ 10$^{15}$-10$^{18}$ gs$^{-1}$ \citep{fra02}.  Hence multicolor-blackbody models cannot account for the secondary maximum flux.  

We fitted multicolor-blackbody models to the underlying outburst flux to check that it could be explained by an optically thick accretion disk.  We chose to examine data at two times:  MJD 52454 and MJD 52487.  Data from the former time was taken from when the primary outburst had passed its peak but still bright relative to quiescent levels.  It was also at this time that we had data in all five bands.  The latter time is that near the peak of the secondary maximum flux.  We constructed the SEDs following the method described here, subtracting the quiescent fluxes of 4U 1543-47 from the outburst fluxes thereby isolating the flux related to the outburst source.  Again, fluxes were converted to mJy using the same zeropoint fluxes mentioned above.  Errors for the SED points for MJD 52454 were calculated from photometric errors.  Errors for the SED of the underlying flux at MJD 52487 were estimated from the error in estimating the baseline connecting points either side of the secondary maximum.  This was found to be $\pm$ 0.1 mag.  

\begin{figure}
\begin{center}
   \plotone{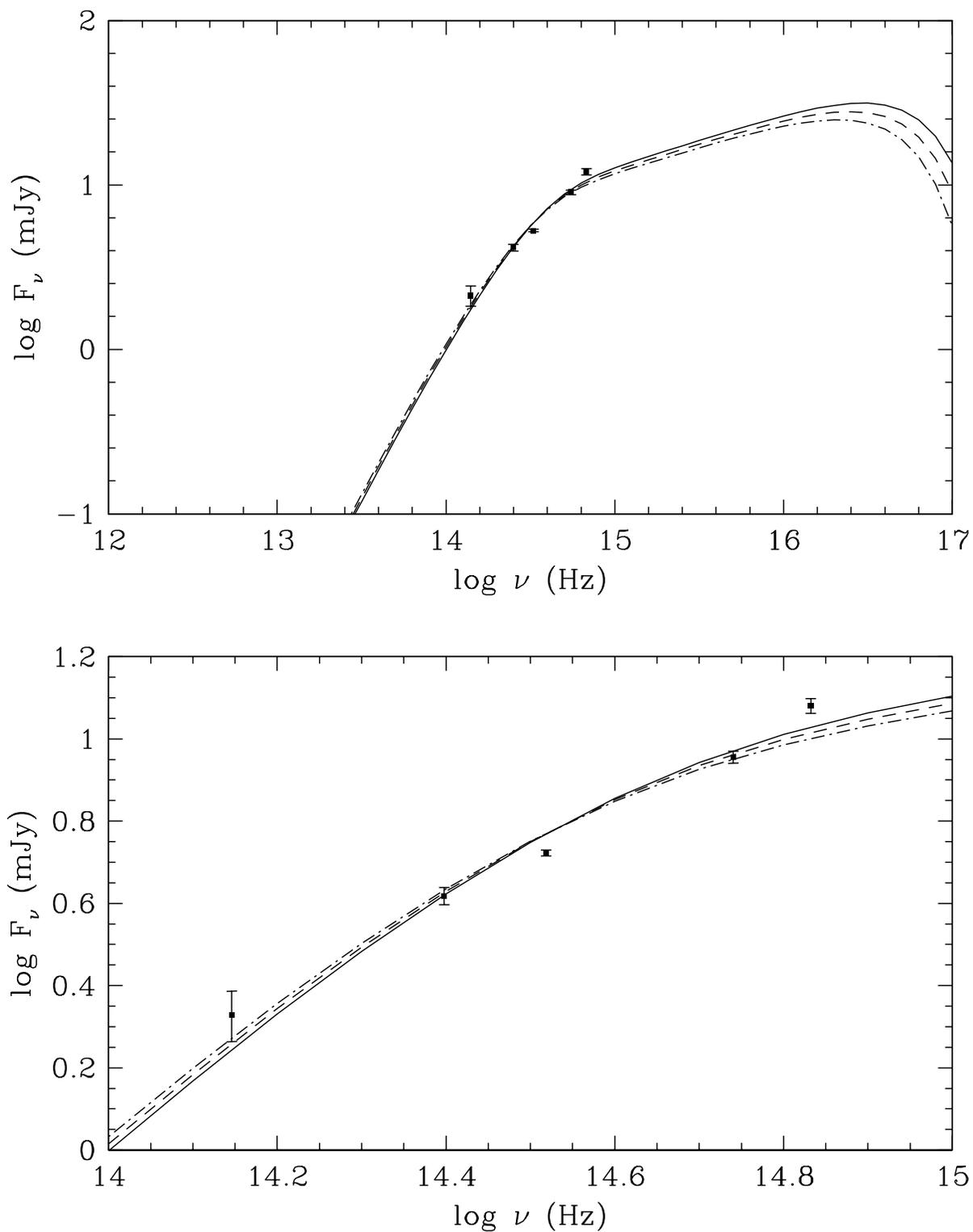}
    \figcaption{Multicolor-blackbody model fits to the SED at MJD 52454 near the primary maximum.  See text for model parameters.  These models fit the data more satisfactorily than that seen for the SED of the secondary maximum flux (Fig. \ref{fig:sed_multi}) and give reasonable values for the mass transfer rate and outer-disk temperature and radius.\label{fig:sed_2454} }
\end{center}
\end{figure}

When constructing multicolor-blackbody models we again fixed r$_{in}$ = 3r$_g$ and M$_1$ = 5 M$_{\odot}$.  Figure \ref{fig:sed_2454} shows the SED at MJD 52454 and three typical multicolor-blackbody fits.  We first chose r$_{out}$ = 10$^5$r$_{in}$ which was within the primary Roche-lobe radius ($\sim$ 5 x 10$^{11}$ cm $\sim$ 10$^6$r$_{in}$).  The resulting fit (shown as solid line) gave $\dot{M}$ = 10$^{19.9}$ gs$^{-1}$ and T$_{out}$ = 5970$^o$K.  This $\dot{M}$ is somewhat higher than usually observed (see above), hence we chose $\dot{M}$ = 10$^{18.0}$ gs$^{-1}$ (dashed line) and 10$^{17.0}$ gs$^{-1}$ (dot-dashed line) for the next two fits.  The corresponding temperatures of the outer disk were 5632$^o$K and 5314$^o$K, and r$_{out}$ = 10$^{4.4}$r$_{in}$ and 10$^{4.1}$r$_{in}$, respectively.  In all three cases the models could provide a good overall match to the SED, although the $\chi^2$ values were never satisfactory, perhaps due to systematic errors from subtracting the flux due to the secondary star.  All temperatures are very reasonable given that the peak of the outburst has passed and that the accretion disk has probably transitioned to a cooler state at this time.  The corresponding $\dot{M}$ are also plausible.

\begin{figure}
\begin{center}
   \plotone{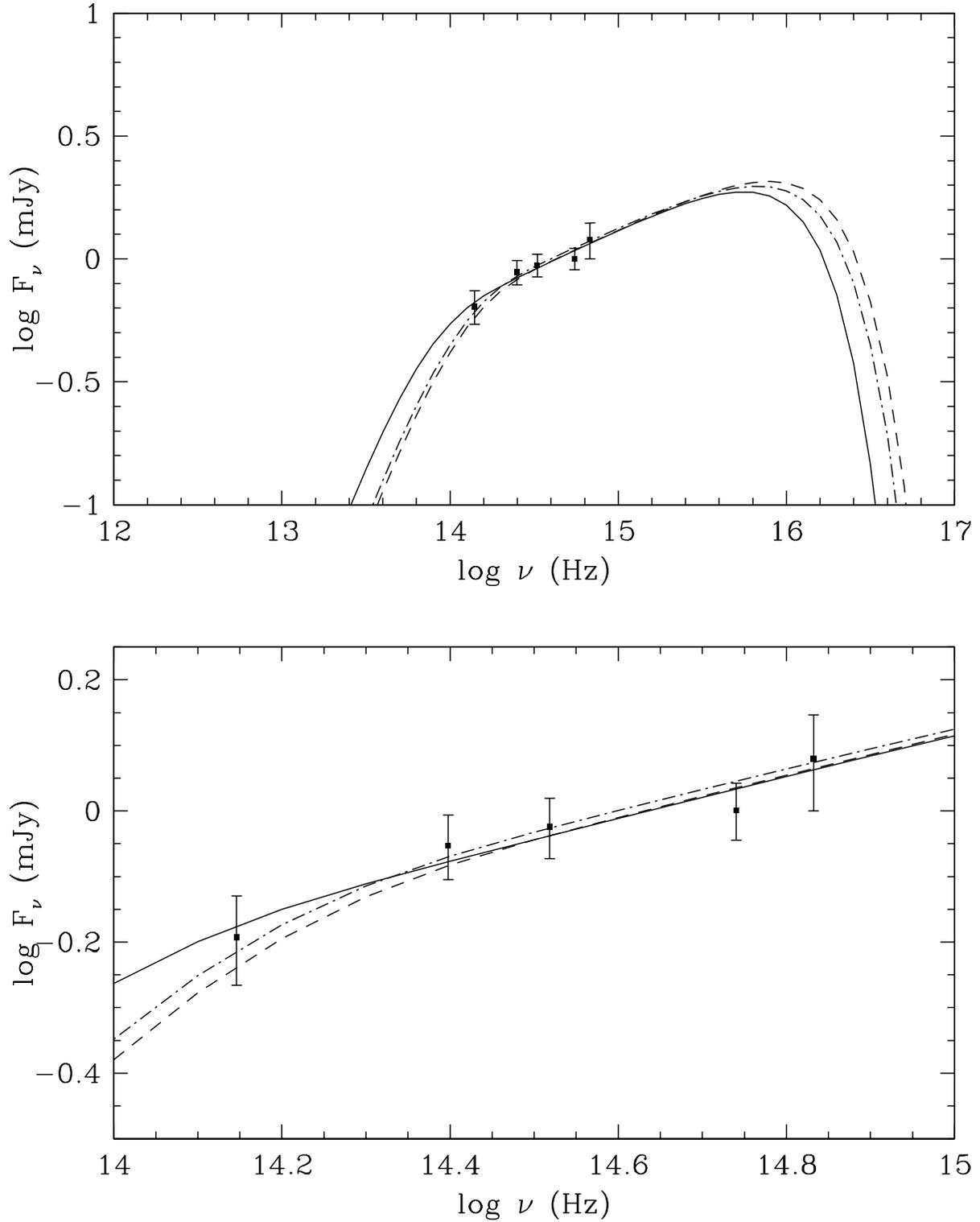}
    \figcaption{Multicolor-blackbody model fits to the SED of the underlying flux of the secondary maximum at MJD 52487.  As for the SED at MJD 52454, these models describe the data well.  See text for model parameters.\label{fig:sed_under} }
\end{center}
\end{figure}

The same fitting procedure was performed for the SED of the underlying flux at MJD 52487, using the same three r$_{out}$ as above.  The SED and resulting fits are shown in Figure \ref{fig:sed_under}.  Again we see good fits with the corresponding temperatures being 1125$^o$K (solid line), 1781$^o$K (dashed line) and 1680$^o$K (dot-dashed line) and  $\dot{M}$ = 10$^{17}$ gs$^{-1}$, 10$^{16}$ gs$^{-1}$ and 10$^{15}$ gs$^{-1}$, respectively.  Although these outer-disk temperatures seem low, fits to ellipsoidal light curves of other systems suggest that T$_{out}$ can be $\le$ 2000$^o$K \citep[e.g.][]{bee02}.

\subsection{Non-thermal emission}

Since the single-blackbody and multicolor-blackbody models cannot adequately describe the SED of the secondary maximum flux, we briefly look at non-thermal emission models. 

\begin{figure}
\begin{center}
   \plotone{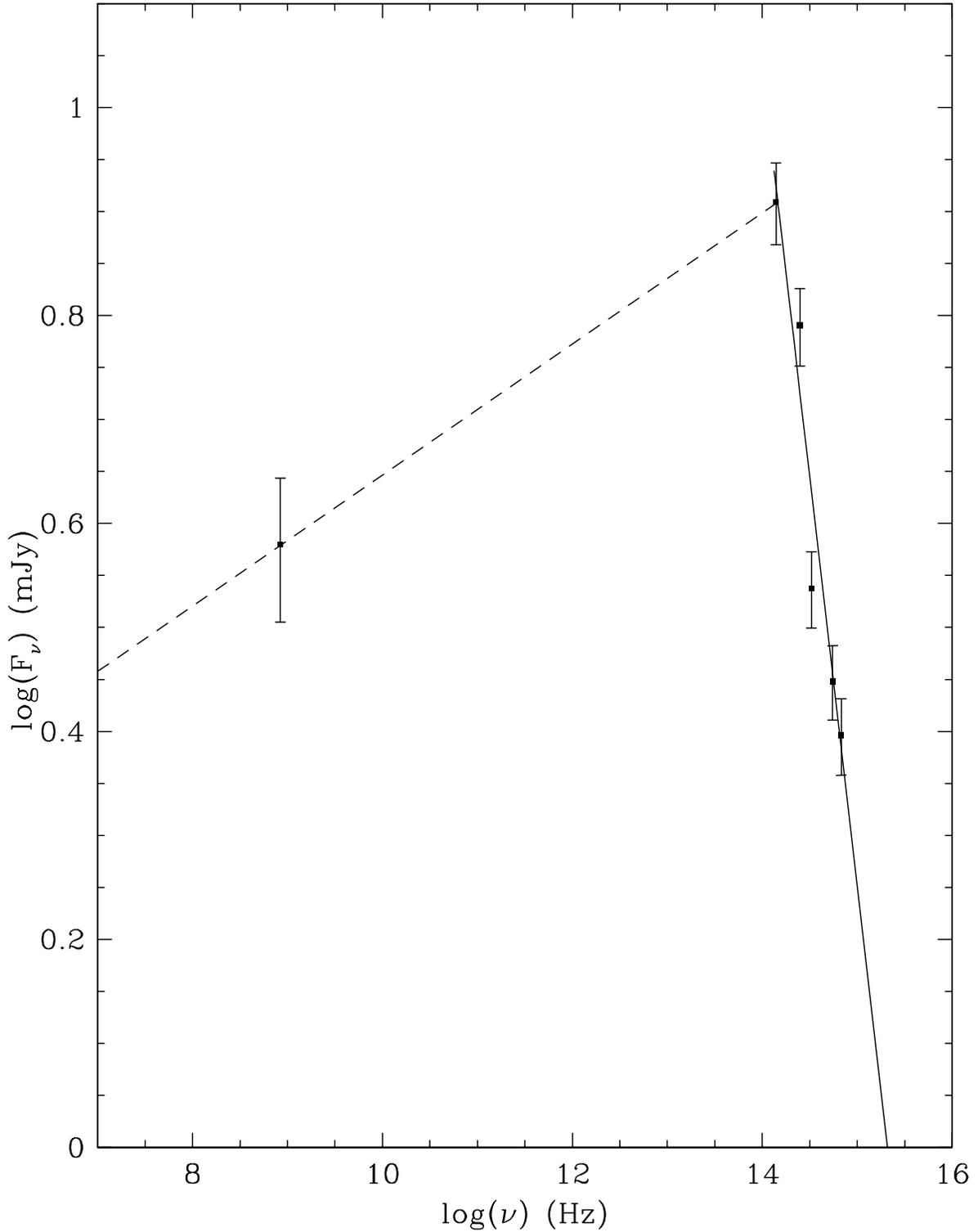}
    \figcaption{\small{Broken power-law fit to the SED.  The solid line passing through the optical and infrared points has a slope of -0.79.  The slope clearly changes between the radio and $K$-band points but we can not say where exactly.  The dashed line joining the radio and $K$-band points has a slope of 0.06 which must be taken as a lower limit.  This is a much better fit than the blackbody models.  Hence, this emission most likely originates from synchrotron radiation.}\label{fig:sed_fit} }
\end{center}
\end{figure}

We fitted the SED of the secondary maximum flux with a broken power-law in
which each piece of the power-law is described by F$_{\nu} \propto \nu^{-\alpha}$.  
We use this model to represent non-thermal emission from a synchrotron source.  The SED and resulting fit are shown in Figure \ref{fig:sed_fit}.  The solid line is the fit to the optical and IR points for which ${\alpha}$ = -0.79 $\pm$ 0.12.  
This power-law fits the optical/IR data well, in contrast to the thermal 
models.
The dashed line is a power-law fit 
passing through the radio point assuming that the break is coincident with the $K$-band point.  
In this case, ${\alpha}$ = 0.063 $\pm$ 0.003.  \textit{This must be taken as a lower limit} as it is possible that the turnover is at longer wavelengths.  The broken power-law provides a much better fit than the thermal, blackbody models.  Hence, we suggest that the source of the secondary maximum emission is synchrotron radiation.  

\section{Discussion}
\label{sec:discussion}

SXTs for which secondary maxima have been observed include A0620-00, GS 1124-68 and GRO J0422+32 \citep{che97} and XTE J1550-564 \citep{jai01}.  In the first three objects, the maxima were observed at short wavelengths (\textit{B} and UV) and were quite weak.  The secondary maximum appeared $\sim$ 40-80 days after the primary X-ray maximum.  The simultaneous soft X-ray light curves, which had not yet reached quiescent levels at the time, also showed secondary maxima.  The soft X-ray light curve of 4U 1543-47 in 1971 \citep{li76} exhibited a secondary maximum $\sim$ 70 days after the primary peak.  No optical nor infrared observations were made at the time.  

XTE J1550-564 \citep{jai01} exhibited very similar behavior to that of 4U 1543-47 presented in this paper.  The secondary maximum appeared $\sim$ 50 days after the the X-ray maximum and increased in strength from optical (\textit{V}- and \textit{I}-band) to infrared (\textit{H}-band) wavelengths.  The ASM had already reached quiescent levels and did not show a flare at the time of the secondary maximum.  There was, however, a deviation in the exponential decay of the RXTE pointed observations during the secondary maximum \citep{tom01}.  \citet{cor01} reported a radio detection near the peak of the secondary maximum.  The detection was made in four bands giving a spectral index of 0.37 $\pm$ 0.10.  

The secondary maxima in these sources have characteristics consistent with the one we have observed in 4U 1543-47.  In particular, the strength of the peak increases from the optical to infrared wavelengths.  The times at which the secondary maximum appears after the primary X-ray peak agree within an order of magnitude.  This timescale could relate to some fundamental parameter of the binary system, such as the orbital period or mass of the compact object.  It seems likely, however, that the trigger of the secondary maxima is related to the transition into the low-hard state.  The secondary maxima of 4U 1543-47 and XTE J1550-564 begin to rise $\sim$ 7 and 8 days, respectively, after the transition into the low-hard state.  Both objects exhibited radio emission associated with a jet which was detected near the peak of the secondary maxima.  Jets are believed to form during the low-hard state \citep{fen01}.  If the secondary maximum emission does in fact originate from jets then one would expect to see it after the object has transitioned into the low-hard state.

\section{Conclusion}
\label{sec:conclusion}

We have presented OIR light curves of the 2002 outburst of 4U 1543-47.  A strong IR flare appears soon after the source transitions into the low-hard state.  Blackbody and broken power-law fits suggest that synchrotron emission is the most likely source of the secondary maximum flux.  The detection of a radio source at the peak of this flare supports this conclusion.  Indeed, the presence of jets during this X-ray state would not be surprising given that we often observe jets in this accretion phase \citep{fen01}.  

Since the secondary maximum is so prominent in the IR, it is clear that IR monitoring of SXTs can successfully trigger other simultaneous observations at other wavelengths and gather useful information into the formation of and flux contribution of jets.  Future observations of such phenomena can be further improved by obtaining simultaneous radio observations in two bands so that a confident fit of the power-law gradient may be obtained.  UV observations will enable us to ascertain any upturn in the SED due to the dominance of an accretion disk blackbody component, although observations in this band are quite difficult due to strong interstellar extinction.  X-ray spectra would also provide valuable information on the relationship to the OIR data and if there is any contribution from thermal IC of disk photons.

\acknowledgments

We would like to thank Emrah Kalemci for his useful comments, Suzanne Tourtellotte and Alfred Chen in assisting with the optical data reduction, and the CTIO observers, David Gonzalez Huerta and Juan Espinoza.  We also thank Darren Depoy for arranging the Stromlo observations.

This paper has made use of the Astrophysics Data System Abstract Service and the ASM/RXTE Data Extraction website.

MB and CB gratefully acknowledge support from the National Science Foundation through grant AST-0098421.

\clearpage


\clearpage 

\clearpage

\begin{table}
\begin{center}
\caption{Two-sided Gaussian parameters for best fits to secondary maxima in $J$- and $K$-band.  Errors shown are 1-$\sigma$.  \label{tab:sec_max_fits}}
\begin{tabular}{lccccc}
\tableline\tableline
Waveband & Amplitude & Peak time & $\sigma_1$ & $\sigma_2$ \\
 & (mag) &  (MJD) & (days) & (days) \\
\tableline
 & & & & \\
$J$ & -1.10$^{+0.15}_{-0.20}$ & 52486.3$^{+1.5}_{-1.8}$ & 1.6$^{+0.8}_{-1.4}$ & 6.1$^{+1.7}_{-2.5}$ \\
 & & & & \\
$K$ & -2.10$^{+0.22}_{-0.23}$ & 52486.5$^{+2.1}_{-1.8}$ & 2.2$^{+1.6}_{-1.6}$ & 6.6$^{+3.9}_{-2.0}$ \\
 & & & & \\
\tableline
\end{tabular}
\end{center}
\end{table}

\end{document}